\documentclass[12pt]{article}
\usepackage{amsfonts,graphicx,amsmath,amsthm,amssymb,xcolor}
\usepackage{epsfig}
\usepackage{float}
\allowdisplaybreaks
\begin{document}

\title{\bf Possible Existence of Ghost Stars in the context of Electromagnetic Field}
\author{Tayyab Naseer$^{1,2}$ \thanks{tayyab.naseer@math.uol.edu.pk; tayyabnaseer48@yahoo.com}~,
K. Hassan$^{1}$ \thanks{komalhassan3@gmail.com}~ and M. Sharif$^{1}$ \thanks{msharif.math@pu.edu.pk}\\
$^1$ Department of Mathematics and Statistics, The University of Lahore,\\
1-KM Defence Road Lahore-54000, Pakistan.\\
$^2$ Research Center of Astrophysics and Cosmology, Khazar University, \\
Baku, AZ1096, 41 Mehseti Street, Azerbaijan.}

\date{}
\maketitle

\begin{abstract}
In this paper, we discuss the existence of ghost star models in the
Einstein-Maxwell framework. In order to explore these objects, we
put forward the idea of Zeldovich and Novikov by keeping in mind
that the energy density of such models lie in the negative range in
some regions of the spacetime geometry. We proceed by taking into
account a static sphere and develop the field equations for a
charged anisotropic fluid configuration. The two generating
functions are then considered and we rewrite the field equations in
terms of the mass and these physical quantities. Afterwards, we
formulate two different models using the conformally flatness
condition along with the considered generating functions. Further,
we adopt the vanishing complexity constraint as well as null active
gravitational mass to find two more solutions. The energy density
for all developed models is also graphically shown. We conclude that
the ghost stars exist in the presence of charge as the energy
density for all the resulting solutions lie in the negative region
for a particular range of the radial coordinate.
\end{abstract}

\section{Introduction}

General Relativity (GR) has become widely accepted theory within the
scientific community as the pre-eminent framework for comprehending
the fundamental nature of gravitational fields. This ground-breaking
theory establishes a direct relationship between the configuration
of matter content characterized by the energy-momentum tensor (EMT)
and the geometrical structure of the spacetime continuum described
by the Einstein tensor, through the renowned field equations
proposed by Albert Einstein in 1915. Scientists have become
increasingly fascinated by the analytical or numerical formulations
of solutions to these equations representing self-gravitating
systems. Schwarzschild's innovative work in calculating their exact
solutions has inspired people working in this field to further
explore their physical attributes, particularly within the framework
of GR its action-based modifications. In 1916, he began with the
premise of a spherically symmetric geometry admitting constant
density and derived different solutions corresponding to both outer
\cite{1} and inner \cite{2} spacetimes.

Researchers have made significant progress in finding solutions to
the gravitational equations which help to model physically realistic
stellar configurations. Nonetheless, calculating these solutions is
a challenging task because these equations contain numerous
geometric terms, resulting in a non-linear system possessing high
degree of freedom. This issue in the equations of motion have
compelled astrophysicists to develop particular techniques to
formulate their solutions so that the physically meaningful results
can be obtained. For instance, the known form of the fluid's energy
density, embedding class-I condition, and the vanishing complexity
are some of the well-known constraints which have extensively been
used in this context. The existing literature presents multiple
solutions in recent times by using these proposed approaches,
describing feasible perfect and anisotropic geometrical structures.
The initial approach often involved coupling the stellar geometry
with an isotropic fluid, where the pressure is assumed to be uniform
in all (tangential and radial) directions. However, the
revolutionary work of Jeans challenged this assumption, suggesting
that multiple factors present within the interior of these systems
may lead to the emergence of anisotropy \cite{13}. This finding
marked a significant shift in the understanding of the complex
dynamics governing self-gravitating objects. Ruderman \cite{14}
conducted a comprehensive analysis and deduced that the celestial
models with the density of at least $10^{15}~g/cm^3$ may exhibit
anisotropic properties. Additionally, it has been established that
certain elements like magnetic field and others surrounding a
compact star can also contribute to the development of anisotropy
\cite{15}-\cite{21d}.

The role of charge in stellar structures is pivotal in understanding
their dynamics and evolution. Charge interactions, primarily
governed by electromagnetic forces, contribute to the stability of
stellar atmospheres and the formation of magnetic fields, which are
crucial in shaping stellar wind and protecting the star from cosmic
radiation. Further, this factor produces a pressure that opposes the
attraction of gravity and thus the geometric structure under
consideration preserves stability. Bekenstein \cite{37fa} confirmed
in his analysis that the repulsive force generated due to the
presence of an electromagnetic field in a dynamical sphere can help
in maintaining the structure's stability. This finding has been
further extended by Esculpi and Aloma \cite{37fb}, who studied the
anisotropic fluid content based geometry and observed that both the
electric charge and positive behavior of anisotropy have a
stabilizing, repulsive effect on the collapsing system. The known
(power-law) form of the charge as a function of the radial
coordinate has been proposed and explored in the analysis of the
physically viable anisotropic stellar configurations coupled with
the electromagnetic field \cite{42aaa,37fc}. Furthermore, Maurya et
al. \cite{37fd} examined the charged inner spacetime geometry
admitting baryonic fluid distribution through the implication of
Karmarkar condition \cite{37fh,37fj} and achieved satisfactory
results. Some recent works in this regard can be found in
\cite{2a}-\cite{2gd}.

By decomposing the Riemann tensor into its orthogonal components,
one can better understand the distribution of mass and the internal
stresses that shape the star's structure. The complexity factor,
derived from this decomposition, serves as a key parameter in
assessing the internal configuration of stellar bodies as Herrera
reported in his recent study \cite{37g}. The motivation behind
proposing such a definition of complexity was the failure of its
previously suggested descriptions. Herrera used the Bel's idea of
decomposition and applied on the anisotropic fluid distribution. He
acquired multiple scalars and found one of them (i.e., $Y_{TF}$) in
accordance with his interest to define the complexity factor. Since
then, this definition has been highly acceptable and utilized by
researchers working in astrophysics. It encapsulates the influence
of various physical factors, including pressure, density, shear etc.
Soon after this, Herrera and his colleagues added the dissipation
flux into the interior fluid and utilized the complexity factor to
study the corresponding patterns of evolution \cite{37h}. Sharif
with his co-researchers \cite{3a}-\cite{3d} extended this definition
for the charged geometries and performed the decomposition on the
Riemann tensor. They concluded that the factor $Y_{TF}$ possesses
the contribution of charge. Many such works have been done in GR and
its action-based modifications from which the acceptability of this
scalar factor has been confirmed \cite{37k}-\cite{37lb}.

From the definition of the mass function in terms of the fluid's
energy density, it becomes clear that the former function must be
non-null in general for positive profile of the later physical
quantity. Thereby, only the acceptance of fluid configurations
possessing negative fluid's density leads to the disappearance of
the total mass. Alternatively, the trivial case where the energy
density takes zero value results in such consequence. Bondi
\cite{1a} raised this point that there could be the fluid
distributions where the energy density becomes negative which was
followed by several researchers. This concept has been discussed in
various scientific contexts, such as classical electron models as
well as the Reissner-Nordstr\"{o}m solution \cite{1b}-\cite{1e}.
Multiple cosmological models have also been constructed by assuming
the fluid distribution admitting negative energy density
\cite{1f,1g}. Additionally, the same scenario has been observed in
the hyperbolic fluids \cite{1h,1i}. However, it is crucial to note
that these examples do not account for quantum effects. Despite
these instances, it is generally accepted that the assumption of
positive density is well-supported at the classical level as it
aligns with our current understanding of the physical world. In the
quantum realm, the dynamics can be quite distinct from the classical
perspective. Recent studies \cite{1j}-\cite{1n} have suggested that
the emergence of negative density is a plausible scenario when
quantum effects become significant. This counterintuitive phenomenon
challenges our conventional understanding of energy and highlights
the unique properties that arise in the quantum domain.

The concept of compact objects with remarkably low total mass
remains viable, provided we embrace the possibility of negative
energy density. These intriguing celestial bodies, which the
astrophysicists referred to ``ghost stars'', draw an analogy to a
somewhat similar phenomenon observed in Einstein-Dirac neutrinos,
known as ghost neutrinos \cite{1o}-\cite{1q}. In this article, we
discuss different solutions to the equations of motion which
represent the ghost stars \cite{1r} within the context of
Einstein-Maxwell scenario. The outline of this work is based on the
following lines. Next section defines the fundamental terms that
help to formulate the field equations for a static spherical
spacetime. We also introduce new fluid variables in terms of the
mass function. Section \textbf{3} considers two generating functions
that are used with particular form of the density to develop two
different models. We then take into account the complexity of
anisotropic fluid source in section \textbf{4} and obtain two novel
solutions. Finally, section \textbf{5} sums up our results and
further provides future directions.

\section{Fundamentals of Charged Spherical Structures}

The Einstein-Hilbert action under the influence of an
electromagnetic field leads to the following equation
\begin{equation}\label{1}
\mathrm{I}=\int \bigg[\frac{R}{8\pi}
+\texttt{L}_{m}+\texttt{L}_{e}\bigg]\sqrt{-g}d^{4}x,
\end{equation}
where determinant of the metric tensor $g_{\phi\chi}$ and
infinitesimal four-volume element are represented by $g$ and
$d^{4}x$, respectively, in the spacetime manifold. Also,
$\texttt{L}_{e}$ is the Lagrangian density of the charge and
$\texttt{L}_{m}$ indicates that of the fluid content within the
interior distribution. The gravitational field equations are
obtained by varying the action \eqref{1} as
\begin{equation}\label{3}
G_{\phi\chi}\equiv
R_{\phi\chi}-\frac{1}{2}Rg_{\phi\chi}=8\pi(T_{\phi\chi}+E_{\phi\chi}).
\end{equation}
Here, $G_{\phi\chi}$ (symbolizes the Einstein tensor) explores the
geometric features, $R_{\phi\chi}$ and $R$ represent the Ricci
tensor and curvature scalar, respectively. Further, $T_{\phi\chi}$
and $E_{\phi\chi}$ characterize the EMT for the usual anisotropic
fluid configuration and electromagnetic field, respectively. Thus,
the EMT decoding the inner matter source is expressed by
\begin{equation}\label{5}
T_{\phi\chi}=\rho
\textsf{V}_{\phi}\textsf{V}_{\chi}+\mathcal{P}h_{\phi\chi}
+\Pi_{\phi\chi},
\end{equation}
where $\rho$, $\textsf{V}_{\phi}$,
$\mathcal{P}=\frac{\mathcal{P}_r+2\mathcal{P}_{\perp}}{3}$ and
$h_{\phi\chi}$ stand for the energy density, four-velocity,
anisotropic pressure ($\mathcal{P}_r\rightarrow$ radial pressure,
$\mathcal{P}_{\perp}\rightarrow$ tangential pressure) and projection
tensor, respectively. Moreover, the quantity $\Pi_{\phi\chi}$ refers
to the anisotropic tensor. They are further expressed as follows
\begin{equation}\label{5a}
h_{\phi\chi}=g_{\phi\chi}-\textsf{V}_{\phi}\textsf{V}_{\chi},
\quad\Pi_{\phi\chi}=\Pi\left(\mathcal{S}_{\phi}\mathcal{S}_{\chi}
+\frac{h_{\phi\chi}}{3}\right),
\quad  \Pi=8\pi(\mathcal{P}_{r}-\mathcal{P}_{\perp}),
\end{equation}
with $\mathcal{S}_{\phi}$ being the four-vector.

The electromagnetic field surrounding the matter sector is
characterized by the EMT \cite{1ra,1rb}
\begin{equation}\label{3b}
E_{\phi\chi}=\frac{1}{4\pi}\left(\mathbb{F}^{n}_{\phi}\mathbb{F}_{\chi
n}-\frac{1}{4}\mathbb{F}_{ab}\mathbb{F}^{ab}g_{\phi\chi}\right),
\end{equation}
where $\mathbb{F}_{\phi\chi}=\varpi_{\chi,\phi}-\varpi_{\phi,\chi}$
denotes the Maxwell field tensor and $\varpi_{\phi}$ with
$\varpi_{\phi}=\varpi(r)\delta^{0}_{\phi}$ indicates the
four-potential. The Maxwell equations in concise form can be written
as
\begin{equation}\nonumber
\mathbb{F}_{[\phi\chi;n]}=0 ,\quad
\mathbb{F}^{\phi\chi}_{~~;\chi}=4\pi \mathbb{J}^{\phi},
\end{equation}
where $\mathbb{J}^{\phi}=\vartheta\textsf{V}^{\phi}$ is called the
four-current vector with $\vartheta$ being the charge density. In
the current setup, these equations turn out to be
\begin{equation}\nonumber
\varpi''+\frac{1}{2r}\{4-(\lambda'+\eta')r\}\varpi'=4\pi\vartheta
e^{\frac{\lambda}{2}+\eta},
\end{equation}
where $'$ symbolizes the partial derivative with respect to $r$.
Using the method of integrating factor on the above equation, we
have
\begin{equation}\nonumber
\varpi'=\frac{q}{r^2}e^{\frac{\lambda+\eta}{2}},
\end{equation}
with $q=\int^{r}_{0}\vartheta e^{\frac{\eta}{2}}\tilde{x}d\tilde{x}$
representing the total charge present inside the fluid
configuration.

The crucial tool that assists in encoding the symmetries of the
spacetime along with the internal modeling of spherical entities, is
defined by the following line element \cite{1rc}
\begin{equation}\label{6}
ds^{2}=e^{\lambda}dt^{2}-e^{\eta}dr^{2}-r^{2}(d\theta^{2}+{\sin^{2}\theta}{d\phi^2}),
\end{equation}
where $\lambda$ and $\eta$ are functions of $r$ only, indicating the
nature of the considered geometry to be static. The quantities
$\textsf{V}^{\phi}$ and $\mathcal{S}^{\phi}$ in view of this metric
are expressed as
\begin{equation}\label{7}
\textsf{V}^{\phi}=\big(e^{\frac{-\lambda(r)}{2}},0,0,0\big),\quad
\mathcal{S}^{\phi}=\big(0,e^{\frac{-\eta(r)}{2}},0,0\big),
\end{equation}
with the distinguished properties
$\textsf{V}_{\phi}\textsf{V}^{\phi}=1$,
$\mathcal{S}_{\phi}\mathcal{S}^{\phi}=-1$ and
$\mathcal{S}^{\phi}\textsf{V}_{\phi}=0$. Equations
\eqref{3}-\eqref{7} produce a set of three field equations within
the framework of electromagnetic field as follows
\begin{align}\label{8}
8\pi
\rho+\frac{q^2}{r^4}&=\frac{1}{r^{2}}+e^{-\eta}\bigg(\frac{\eta'}{r}-\frac{1}{r^{2}}\bigg),
\\\label{9}
8\pi \mathcal{P}_r-\frac{q^2}{r^4}&
=-\frac{1}{r^{2}}+e^{-\eta}\bigg(\frac{1}{r^{2}}+\frac{\lambda'}{r}\bigg),
\\\label{10}
8\pi \mathcal{P}_{\perp}+\frac{q^2}{
r^4}&=\frac{e^{-\eta}}{4}\bigg(2\lambda''+\lambda'^{2}-\lambda'\eta'+\frac{2\lambda'}{r}
-\frac{2\eta'}{r}\bigg).
\end{align}

Now, we look at the general algorithm of finding solution to the
anisotropic system with the help of two generating functions.
Following this, we take Eqs.\eqref{9} and \eqref{10}
\begin{align}\nonumber
8\pi(\mathcal{P}_{r}-\mathcal{P}_{\perp})
&=e^{-\eta}\bigg[\frac{\lambda'}{2r}-\frac{\lambda''}{2}
-\bigg(\frac{\lambda'}{2}\bigg)^2+\frac{1}{r^2}
+\frac{\eta'}{2}\bigg(\frac{\lambda'}{2}+\frac{1}{r}\bigg)\bigg]
-\frac{1}{r^2}-\frac{2q^2}{r^4}.\\\label{c67}
\end{align}
It can also be observed that the above system consists of three
differential equations with six unknowns
($\lambda,\eta,\rho,\mathcal{P}_r,\mathcal{P}_{\perp},q$). To close
this system, we require three more constraints. Thus, we introduce
two metric coefficients of the form \cite{1rd}
\begin{equation}\label{14}
e^{\lambda}=e^{\int(2z-\frac{2}{r})dr},\quad e^{-\eta}=w(r),
\end{equation}
and the substitution of these metric functions in \eqref{c67}
produces the following differential equation
\begin{align}\label{67}
w'+w\bigg[\frac{4}{r^2z}+\frac{2z'}{z}-\frac{6}{r}+2z\bigg]&=
-\frac{2}{z}\bigg[\frac{1}{r^2}-\frac{2q^2}{r^4} +\Pi\bigg],
\end{align}
whose integration yields the $g_{rr}^{-1}$ component in terms of
generating functions given by
\begin{equation}\label{62}
e^{-\eta}=\frac{z^2e^{\int
\big(\frac{4}{r^2z}+2z\big)dr}}{r^6\bigg\{\textsf{C}-2\int\frac{ze^{\int
\big(\frac{4}{r^2z}+2z\big)dr}(1+\Pi
r^2-\frac{2q^2}{r^2})}{r^8}dr\bigg\}},
\end{equation}
where $\textsf{C}$ is the integration constant. Moreover,
Eqs.\eqref{14} and \eqref{62} together compose the following line
element
\begin{align}\nonumber
ds^2&=e^{\int(2z-2/r)dr}dt^2-\frac{z^2e^{\int
\big(\frac{4}{r^2z}+2z\big)dr}}{r^6\bigg\{\textsf{C}-2\int\frac{ze^{\int
\big(\frac{4}{r^2z}+2z\big)dr}(1+\Pi
r^2-\frac{2q^2}{r^2})}{r^8}dr\bigg\}}dr^2\\\label{c70}
&-r^2(d\theta^2+\sin^2\theta d\phi^2).
\end{align}
It is worthy to notice here that the spherically symmetric
anisotropic distributions can be thoroughly explained via generating
functions. Thus, to achieve a physically relevant and meaningful
model, it is convenient to rearrange the field equations in the form
of generating functions and metric potentials. Hence the transformed
field equations are presented as follows
\begin{align}\label{15}
4\pi\rho&=\frac{m'}{r^2}-\frac{qq'}{r^3},\\\label{16}
4\pi\mathcal{P}_{r}&=\frac{\frac{m}{r}+z(r-2m)-1}{r^2}+\frac{q^2z}{r^3},\\\nonumber
4\pi\mathcal{P}_{\perp}&=\bigg(\frac{q^2}{r^2}-\frac{2m}{r}+1\bigg)\bigg(z^2+z'
+\frac{1}{r^2}-\frac{z}{r}\bigg)\\\label{17}&+z\bigg(\frac{qq'}{r^2}-\frac{q^2}{r^3}
-\frac{m'}{r}+\frac{m}{r^2}\bigg)-\frac{q^2}{r^4},
\end{align}
where the mass function $m$ in the presence of charge is defined as
\cite{1x}
\begin{equation}\label{63}
e^{-\eta}=1-\frac{2m}{r}+\frac{q^2}{r^2}.
\end{equation}

Darmois junction conditions play a crucial role in matching the
interior and exterior structures over the hypersurface $\Sigma$. The
interior geometry is adopted in Eq.\eqref{6}, admitting the effect
of the charge, thus the outer regime is considered by the
Reissner-Nordstr\"{o}m metric as
\begin{equation}\label{21}
ds^2=\bigg(1-\frac{2\mathcal{M}}{r}+\frac{\mathcal{Q}^2}{r^2}\bigg)dt^2
-\bigg(1-\frac{2\mathcal{M}}{r}+\frac{\mathcal{Q}^2}{r^2}\bigg)^{-1}dr^2
-r^{2}(d\theta^{2}+{\sin^{2}\theta}{d\phi^2}),
\end{equation}
where $\mathcal{M}$ and $\mathcal{Q}$ represent the respective mass
and charge, respectively. The continuity of the first and second
fundamental forms evaluates the following conditions over the
spherical interface as
\begin{align}\label{18}
e^{\lambda}&~{_=^\Sigma}~1-\frac{2\mathcal{M}}{r_{\Sigma}}
+\frac{\mathcal{Q}^2}{r_{\Sigma}^2},\\\label{19}
e^{-\eta}&~{_=^\Sigma}~1-\frac{2\mathcal{M}}{r_{\Sigma}}
+\frac{\mathcal{Q}^2}{r_{\Sigma}^2},\\\label{20}
\mathcal{P}_{r}&~{_=^\Sigma}~0, \quad \mathcal{Q}~{_=^\Sigma}~q.
\end{align}
It is interesting to note that the above constraints are satisfied
for all values of $\mathcal{M}$ (including $\mathcal{M}=0$). If we
assume the total mass to be null, Eqs.\eqref{16} and \eqref{20}
together derive the following result
\begin{equation}\label{22}
z_{\Sigma}=\frac{r_{\Sigma}}{r_{\Sigma}^2+\mathcal{Q}^2}.
\end{equation}
A family of different constraints shall be taken in the following
sections to explore the existence of spherically symmetric ghost
star configurations. It should be mentioned here that some of these
constraints have already been discussed in the literature while
studying relativistic compact stars under certain physical
conditions.

\section{Conformally Flat Ghost Stars}

There exist only one independent component of the Weyl tensor in the
case of a 4-dimensional spherically symmetric spacetime. If this
geometry is considered to be the conformally flat, we are left with
the differential equation after nullifying that component. This
equation leads to the following generating function \cite{40}
\begin{align}\label{23}
z&=\frac{2}{r}\pm\frac{e^{\frac{\eta}{2}}}{r}
\tanh\bigg(\int\frac{e^{\frac{\eta}{2}}}{r}dr\bigg).
\end{align}
Merging the same constraint with Eq.\eqref{17}, we get the following
condition
\begin{align}\label{24}
\Pi&=\bigg(\frac{1-e^{-\eta}}{r^2}\bigg)'r.
\end{align}
It needs to be mentioned here that we impose the negative sign in
Eq.\eqref{23} in order to construct an acceptable model. In the
following subsections, we implement the conformal flatness
constraint together with some more conditions to develop two charged
ghost star models.

\subsection{Model I}

To proceed, we assume an expression of the energy density as follows
\begin{equation}\label{24a}
4\pi\rho=\sum^{n}_{h=0}b_{h}r^{h-2},
\end{equation}
which is used in Eq.\eqref{15} to evaluate the value of mass after
integration as
\begin{equation}\label{26}
m=\sum^{n}_{h=0}\frac{b_{h}r^{h+1}}{h+1}+\frac{3\beta^2r^5}{5},
\end{equation}
where we have used $q=\beta r^{3}$ with $\beta$ having the dimension
of $\frac{1}{\ell^2}$ \cite{42aaa}. In ghost stars, we already
mentioned that the mass at the boundary vanishes so the following
condition must be satisfied
\begin{equation}\label{27}
\sum^{n}_{h=0}\frac{b_{h}r_{\Sigma}^{h+1}}{h+1}+\frac{3\beta^2r_{\Sigma}^5}{5}=0.
\end{equation}
As in Eq.\eqref{24a}, $n$ can be any finite number, therefore, in
the current setup, we restrict its value up to 2 to describe a
particular model. Under this restriction, the energy density
\eqref{24a} and mass function \eqref{26} are expressed through the
following expressions
\begin{align}\label{28}
4\pi\rho&=\frac{b_1}{r}-\frac{3}{2r^2}+b_2,\\\label{28a}
m&=\frac{b_1 r^2}{2}+\frac{b_2
r^3}{3}-\frac{3r}{2}+\frac{3\beta^2r^5}{5},
\end{align}
where $b_0,b_1$ and $b_2$ are the constant terms corresponding to
$h=0,1$ and 2, respectively. Moreover, we have assumed
$b_0=-\frac{3}{2}$ for the sake of simplicity. Using this in
Eq.\eqref{27}, we extract the value of $b_2$ as
\begin{equation}\label{29}
b_2=\frac{9}{2r_{\Sigma}^2}-\frac{3b_1}{2r_{\Sigma}}-\frac{9\beta
r_{\Sigma}^2}{5}.
\end{equation}
Substituting Eqs.\eqref{28a} and \eqref{29} in \eqref{63}, we obtain
the metric coefficient as
\begin{equation}\label{30}
e^{-\eta}=b_1r\bigg(\frac{r}{r_{\Sigma}}-1\bigg)-\frac{3r^2}{r_{\Sigma}^{2}}+\frac{6\beta
r^2}{5}(r_{\Sigma}^{2}-r^2)+4.
\end{equation}

The above metric function helps to formulate the two generating
functions \eqref{23} and \eqref{24} of the form
\begin{align}\nonumber
z&=\frac{2}{r}- \big(5 b_1 r+\beta r^2-\tau_1-12 r_{\Sigma} ^2 \beta
-40\big)\bigg\{r \big(12 r_{\Sigma} ^2 \beta -5 b_1 r+\beta
r^2+\tau_1 +40\big)\\\label{31}&\sqrt{\bigg(\frac{5 b_1 r
(r-r_{\Sigma} )}{r_{\Sigma} }+6 \beta  r^2
\big(r_{\Sigma}^2-r^2\big)-\frac{15 r^2}{r_{\Sigma} ^2}+20\bigg)}
\bigg\}^{-1},\\\label{32} \Pi&=\frac{-5 b_1 r+12 \beta  r^4+30}{5
r^2},
\end{align}
where $\tau_1$ is given by
$$\tau_1=2 \sqrt{6 r_{\Sigma} ^2 \beta
+20} \sqrt{6 r_{\Sigma} ^2 \beta +\frac{5 b_1 r (r-r_{\Sigma}
)}{r_{\Sigma} }-r^2 \bigg(\frac{15}{r_{\Sigma} ^2}+6 \beta
\bigg)+20}.$$
Here, using the condition \eqref{22} in \eqref{31}, we
can easily compute the value of $b_1$ as
\begin{align}\nonumber
b_1&=\frac{1}{5 \big(\sqrt{5} r_{\Sigma} ^5 \beta ^2+2 r_{\Sigma} ^5
\beta ^2+\sqrt{5} r_{\Sigma} +r_{\Sigma} \big)}\big\{11 \sqrt{5}
r_{\Sigma} ^6 \beta ^3+40 \sqrt{5} r_{\Sigma} ^4 \beta
^2\\\nonumber&+11 \sqrt{5} r_{\Sigma} ^2 \beta  +2 \sqrt{10} \sqrt{3
r_{\Sigma} ^2 \beta +10}+10 \sqrt{2} \sqrt{3 r_{\Sigma} ^2 \beta
+10}+26 r_{\Sigma} ^6 \beta ^3\\\nonumber&+4 \sqrt{10} r_{\Sigma} ^4
\beta ^2 \sqrt{3 r_{\Sigma} ^2 \beta +10}+10 \sqrt{2} r_{\Sigma} ^4
\beta ^2 \sqrt{3 r_{\Sigma} ^2 \beta +10}+80 r_{\Sigma} ^4 \beta
^2\\\label{33}&+13 r_{\Sigma} ^2 \beta+40 \sqrt{5}+40\big\},
\end{align}
and substituting this back into Eq.\eqref{29} gives the value of
$b_2$ expressed by
\begin{align}\nonumber
b_2&=\frac{3}{10} \bigg[\beta \bigg\{\frac{2 \sqrt{5}}{\big(9+4
\sqrt{5}\big) r_{\Sigma} ^4 \beta ^2+3 \sqrt{5}+7}-4
\sqrt{5}-3\bigg\}
\\\label{33a}&-6 r_{\Sigma} ^2 \beta-\frac{2 \sqrt{30 r_{\Sigma} ^2 \beta +100}+25}{r_{\Sigma}
^2}\bigg].
\end{align}
With the help of the above two constants, we get the final
expressions of $z$ and $m$ using \eqref{28a} and \eqref{31} as
\begin{align}\nonumber
z&=\frac{2}{r}-5 \big(\beta r^2-\tau_4-12 r_{\Sigma} ^2 \beta
+\frac{r \tau _2}{r_{\Sigma}  \tau _3}-40\big)\bigg\{r\big(12
r_{\Sigma} ^2 \beta +\tau_4+\beta r^2\\\label{34}&-\frac{r \tau
_2}{r_{\Sigma} \tau _3}+40\big)\sqrt{\big(6 \beta r^2
\big(r_{\Sigma} ^2-r^2\big)-\frac{15 r^2}{r_{\Sigma} ^2}+\frac{r
\tau _2 (r-r_{\Sigma} )}{r_{\Sigma} ^2 \tau _3}+20\big)}
\bigg\}^{-1},\\\nonumber m&=\frac{1}{10} r \bigg[6 \beta ^2 r^4+r^2
\bigg\{\beta  \bigg(\frac{2 \sqrt{5}}{\big(9+4 \sqrt{5}\big)
r_{\Sigma} ^4 \beta ^2+3 \sqrt{5}+7}-4 \sqrt{5}-3\bigg)
\\\nonumber&-6 r_{\Sigma}
^2 \beta+\frac{1}{r_{\Sigma} ^2}\{-2 \sqrt{30 r_{\Sigma} ^2 \beta
+100}-25\}\bigg\}+r \big\{r_{\Sigma} ^2 \beta  \big(r_{\Sigma} ^2
\beta \big(r_{\Sigma} ^2\beta\big(26
\\\nonumber&+11 \sqrt{5}\big)+10 \sqrt{6 r_{\Sigma} ^2 \beta +20}+4 \sqrt{30
r_{\Sigma} ^2 \beta +100}+40 \sqrt{5}+80\big)\\\nonumber&+11
\sqrt{5}+13\big)+2 \big(5 \sqrt{6 r_{\Sigma} ^2 \beta +20}+\sqrt{30
r_{\Sigma} ^2 \beta +100}+20
\sqrt{5}+20\big)\big\}\\\label{34a}&\times \{r_{\Sigma}
\big(\sqrt{5}+\big(2+\sqrt{5}\big) r_{\Sigma} ^4 \beta
^2+1\big)\}^{-1}-15\bigg],
\end{align}
where
\begin{align}\nonumber
\tau _2&=\big(26+11 \sqrt{5}\big) r_{\Sigma} ^6 \beta ^3+\big(13+11
\sqrt{5}\big) r_{\Sigma} ^2 \beta +2 \big(5 \sqrt{6 r_{\Sigma} ^2
\beta +20}+20 \sqrt{5}\\\nonumber&+20+\sqrt{30 r_{\Sigma} ^2 \beta
+100}\big)+2 r_{\Sigma} ^4 \beta ^2 \big(5 \sqrt{6 r_{\Sigma} ^2
\beta +20}+2 \sqrt{30 r_{\Sigma} ^2 \beta +100}\\\nonumber&+20
\sqrt{5}+40\big),\\\nonumber \tau _3&=\big(2+\sqrt{5}\big)
r_{\Sigma} ^4 \beta ^2+\sqrt{5}+1,\\\nonumber\tau_4&=2 \sqrt{6
r_{\Sigma} ^2 \beta +20} \sqrt{6 r_{\Sigma} ^2 \beta -r^2
\bigg(\frac{15}{r_{\Sigma} ^2}+6 \beta \bigg)+\frac{r \tau _2
(r-r_{\Sigma} )}{r_{\Sigma} ^2 \tau _3}+20}.
\end{align}
Also, the final expressions of $\rho$ and $\Pi$ are evaluated by
inserting the values of $b_1$ and $b_2$ in \eqref{28} and
\eqref{32}, respectively
\begin{align}\nonumber
4\pi\rho&=\frac{3}{2 r^2}+\frac{3}{10} \bigg\{\beta  \bigg(\frac{2
\sqrt{5}}{\big(9+4 \sqrt{5}\big) r_{\Sigma} ^4 \beta ^2+3
\sqrt{5}+7}-4 \sqrt{5}-3\bigg)-6 r_{\Sigma} ^2 \beta
\\\nonumber&+\frac{-2 \sqrt{30 r_{\Sigma} ^2 \beta +100}-25}{r_{\Sigma}
^2}\bigg\}+\frac{1}{5 r_{\Sigma} r \big\{\big(2+\sqrt{5}\big)
r_{\Sigma} ^4 \beta^2+\sqrt{5}+1\big\}}\big\{\big(26\\\nonumber&+11
\sqrt{5}\big) r_{\Sigma} ^6 \beta ^3+\big(13+11 \sqrt{5}\big)
r_{\Sigma} ^2 \beta +2 \big(5 \sqrt{6 r_{\Sigma} ^2 \beta +20}+20
\sqrt{5}+20\\\nonumber&+\sqrt{30 r_{\Sigma} ^2 \beta +100}\big)+2
r_{\Sigma} ^4 \beta ^2 \big(5 \sqrt{6 r_{\Sigma} ^2 \beta +20}+2
\sqrt{30 r_{\Sigma} ^2 \beta +100}\\\label{25}&+20
\sqrt{5}+40\big)\big\},\\\nonumber \Pi&=\big[30 r_{\Sigma}
\big\{\big(2+\sqrt{5}\big) r_{\Sigma} ^4 \beta
^2+\sqrt{5}+1\big\}+12 r_{\Sigma}  \beta  r^4
\big\{\big(2+\sqrt{5}\big) r_{\Sigma} ^4 \beta
^2+\sqrt{5}\\\nonumber&+1\big\}-r \big\{\big(26+11 \sqrt{5}\big)
r_{\Sigma} ^6 \beta ^3+\big(13+11 \sqrt{5}\big) r_{\Sigma} ^2 \beta
+2 \big(5 \sqrt{6 r_{\Sigma} ^2 \beta +20}\\\nonumber&+20
\sqrt{5}+20+\sqrt{30 r_{\Sigma} ^2 \beta +100}\big)+ \big(5 \sqrt{6
r_{\Sigma} ^2 \beta +20}+2 \sqrt{30 r_{\Sigma} ^2 \beta
+100}\\\label{25a}&+20 \sqrt{5}+40\big)2 r_{\Sigma} ^4 \beta
^2\big\}\big]\big[5 r_{\Sigma} r^2 \big\{\big(2+\sqrt{5}\big)
r_{\Sigma} ^4 \beta ^2+\sqrt{5}+1\big\}\big]^{-1}.
\end{align}
Due to lengthy expressions of $\mathcal{P}_r$ and
$\mathcal{P}_{\perp}$, we do not mention their values. It is much
interesting to know that the conformal flatness condition is
preserved because the following condition holds when using
\eqref{25} and \eqref{25a} as
\begin{equation}\label{61}
\mathcal{P}_r-\mathcal{P}_{\perp}=\frac{1}{r^3}\int^{r}_{0}r^3\rho'dr-\frac{3\beta^2
r^2}{40\pi}.
\end{equation}

From Figure \textbf{1}, we notice that the energy density \eqref{25}
becomes negative in some spacetime regions for different choices of
the free parameter $\beta$ and $r_{\Sigma}=5$. This guarantees our
model to be a suitable candidate of a ghost star. The highlights are
given as follows.
\begin{itemize}
\item For $\beta=-0.02$, we find the energy density to be negative for
$r<0.634$.
\item For $\beta=-0.04$, this becomes negative for $r<0.575$.
\item For $\beta=-0.06$, negative profile of the energy density is
observed for $r<0.544$.
\item For $\beta=-0.08$, this fluid parameter is noticed to be
negative for $r<0.516$.
\end{itemize}
\begin{figure}\center
\epsfig{file=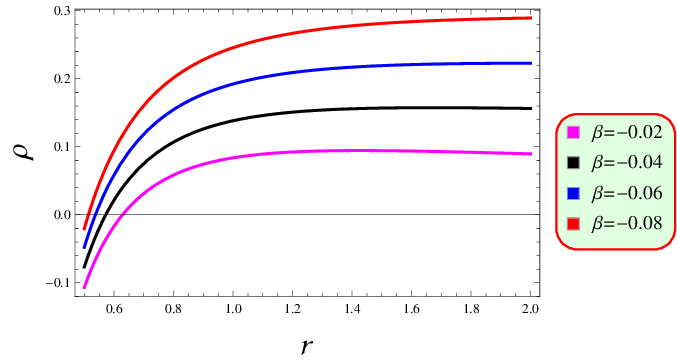,width=0.5\linewidth} \caption{Energy density
\eqref{25} versus $r$ for model I.}
\end{figure}

\subsection{Model II}

Gokhroo and Mehra \cite{41} suggested a specific approach with
varying energy density profile to achieve an acceptable model for
compact stars. We now assume the metric function of the following
form \cite{1re}
\begin{equation}\label{36}
e^{-\eta}=1-\gamma r^2+\frac{3\textsf{k}\gamma
r^4}{5r^{2}_{\Sigma}}.
\end{equation}
Utilizing this metric component in \eqref{8} and \eqref{14} leads to
the following forms of the energy density and mass function,
respectively, as
\begin{align}\label{37}
\rho&=\rho_{o}\bigg(1-\frac{\textsf{k}r^2}{r^{2}_{\Sigma}}\bigg)-\frac{\beta^2r^2}{8\pi},\\\label{38}
m&=\frac{\beta^2
r^5}{2}+\frac{4\pi\rho_{o}r^3}{3}\bigg(1-\frac{3\textsf{k}r^2}{5r^{2}_{\Sigma}}\bigg),
\end{align}
where $\rho_{o}=\frac{3\gamma}{8\pi}$. For $m(r_{\Sigma})=0$, we
obtain
$\textsf{k}=\frac{5\beta^2r^2_{\Sigma}}{3\gamma}+\frac{5}{3}$. Now,
we replace this value of $\textsf{k}$ in \eqref{36}-\eqref{38} and
obtain the metric function, density profile and mass function as
\begin{align}\label{38a}
e^{-\eta}&=\frac{12\bigg(\frac{5\beta^2r^2_{\Sigma}}{12}
+\frac{5}{3}\bigg)r^4}{5r^{4}_{\Sigma}}-\frac{4r^2}{r^{2}_{\Sigma}}+1,\\\label{39}
4\pi\rho&=\frac{6}{r^{2}_{\Sigma}}\bigg(1-\frac{5\beta^2r^2_{\Sigma}r^2}{12}
+\frac{5r^2}{3r^{2}_{\Sigma}}\bigg)-2\beta^2r^2,\\\label{40}
m&=\frac{\beta^2
r^5}{2}+\frac{2r^3}{r^{2}_{\Sigma}}\bigg(1-\frac{\beta^2r^2_{\Sigma}r^2}{4}
-\frac{r^2}{r^{2}_{\Sigma}}\bigg),
\end{align}
where we restrict the value of $\gamma$ to be $\frac{4}{r_{\Sigma}}$
in order to ease the calculations. Thus, utilizing the metric
potential \eqref{38a}, we get the generating functions under this
ansatz as
\begin{align}\label{41}
z&=\frac{2 r_{\Sigma} ^2 \sqrt{\beta^4 r^4+\big(r_{\Sigma}^2-2
r^2\big)^2}-r_{\Sigma}^2 \big(\beta^4+1\big) r^2}{r \sqrt{\beta^4
r^4+\big(r_{\Sigma}^2-2 r^2\big)^2} \big(2
r_{\Sigma}^2-\big(\beta^4+3\big) r^2\big)}+\frac{2}{r}
,\\\label{42}\Pi&=-\frac{2 \big(\beta ^4+4\big) r^2}{r_{\Sigma}^4}.
\end{align}
Moreover, the final expressions of $m$ and $z$ are used in the field
equations \eqref{16} and \eqref{17} to compute the expressions of
$\mathcal{P}_r$ and $\mathcal{P}_{\perp}$. They are expresses as
\begin{align}\nonumber
4\pi \mathcal{P}_r&=\beta ^2 r^3 \tau_5 +\frac{1}{2 r_{\Sigma} ^6
r^2}\bigg[4 r_{\Sigma} ^4 r^2-2 r_{\Sigma} ^6-\beta ^2
r^{10}+r_{\Sigma} ^2 r^4 \big(r_{\Sigma} ^4 \beta
^2-4\big)\\\label{43}&+2 r_{\Sigma} ^6 \tau_5 \bigg\{\frac{\beta ^2
r^{11}}{r_{\Sigma} ^6}+r^5 \bigg(\frac{4}{r_{\Sigma} ^4}-\beta
^2\bigg)-\frac{4 r^3}{r_{\Sigma} ^2}+r\bigg\}\bigg],\\\nonumber 4\pi
\mathcal{P}_{\perp}&=\frac{1}{r_{\Sigma} ^6 \big\{\big(\beta
^4+3\big) r^3-2 r_{\Sigma} ^2 r\big\}^2 \big\{r_{\Sigma}
^4+\big(\beta ^4+4\big) r^4-4 r_{\Sigma} ^2
r^2\big\}^{3/2}}\big[\big(r_{\Sigma} ^6\\\nonumber&+\beta ^2
r^{10}+4 r_{\Sigma} ^2 r^4-4 r_{\Sigma} ^4 r^2\big) \big\{16
r_{\Sigma} ^8 \tau_6 +\big(\beta ^4+3\big)^2 \big(\beta ^4+4\big)
r^8 \tau_6
\\\nonumber&-4 r_{\Sigma} ^2 r^6 \big(r_{\Sigma} ^2 \big(3 \beta
^8+14 \beta ^4+11\big)+\big(2 \beta ^8+13 \beta ^4+21\big) \tau_6
\big)+2 r_{\Sigma} ^4 r^4\\\nonumber&\times \big(r_{\Sigma} ^2
\big(\beta ^8+24 \beta ^4+23\big)+\big(\beta ^8+20 \beta ^4+61\big)
\tau_6 \big)-4 r_{\Sigma} ^6 r^2 \big(3 r_{\Sigma} ^2 \big(\beta
^4\\\nonumber&+1\big)+\big(\beta ^4+19\big) \tau_6
\big)\big\}\big]-\frac{1}{r_{\Sigma} ^6 \tau_6 \big\{2 r_{\Sigma}
^2-\big(\beta ^4+3\big) r^2\big\}}\big[\big(5 \beta ^2
r^8\\\label{44}&-4 r_{\Sigma} ^4+8 r_{\Sigma} ^2 r^2\big) \big\{r^2
\big(r_{\Sigma} ^2 \big(\beta ^4+1\big)+2 \big(\beta ^4+3\big)
\tau_6 \big)-6 r_{\Sigma} ^2 \tau_6 \big\}\big]-\frac{1}{2} \beta ^2
r^2,
\end{align}
where
\begin{align}\nonumber
\tau_5&=\frac{2 \big(\big(\beta ^4+3\big) r^2-3 r_{\Sigma}
^2\big)}{r \big(\big(\beta ^4+3\big) r^2-2 r_{\Sigma}
^2\big)}+\frac{r_{\Sigma} ^2 \big(\beta ^4+1\big) r}{\big(\big(\beta
^4+3\big) r^2-2 r_{\Sigma} ^2\big) \sqrt{r_{\Sigma} ^4+\big(\beta
^4+4\big) r^4-4 r_{\Sigma} ^2 r^2}},\\\nonumber
\tau_6&=\sqrt{r_{\Sigma}^4+\left(\beta ^4+4\right) r^4-4r_{\Sigma}^2
r^2}.
\end{align}
\begin{figure}\center
\epsfig{file=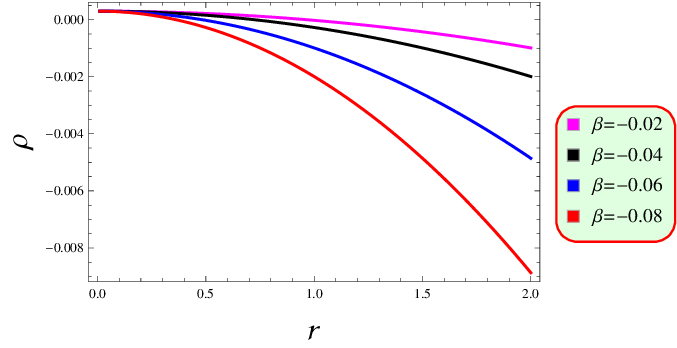,width=0.5\linewidth} \caption{Energy density
\eqref{39} versus $r$ for model II.}
\end{figure}

Figure \textbf{2} depicts the negative profile of the energy density
\eqref{39} for different values of $\beta$. The detailed analysis is
given as follows.
\begin{itemize}
\item For $\beta=-0.02$, negative profile of the energy density is
observed for $r>0.956$.
\item For $\beta=-0.04$, this becomes negative for $r>0.718$.
\item For $\beta=-0.06$, we find the energy density to be negative for
$r>0.452$.
\item For $\beta=-0.08$, this fluid parameter is noticed to be
negative for $r>0.339$.
\end{itemize}

\subsection{Sensitivity Analysis of Energy Density
Functions}

It is important to clarify that while our models utilize particular
energy density profiles \eqref{24a} and \eqref{37}, the underlying
physics is rooted in the robust framework of GR and electromagnetic
theory. The Einstein-Maxwell equations allow for a variety of energy
density configurations, and our exploration of ghost stars is not
limited to a single formulation. In fact, we have considered
multiple configurations, and our findings indicate that the
essential characteristics of ghost stars - specifically, their
negative energy density in certain regions - are consistent across
different models. This suggests that our conclusions are not
artifacts of any one model but rather reflect a broader physical
reality. To further substantiate this point, we conduct numerical
simulations that explore a wider range of energy density. By varying
charge parameter and examining its effects, we demonstrate that the
existence of ghost stars is a robust phenomenon that emerges under
various conditions. These simulations may provide additional
evidence that supports our theoretical predictions and illustrates
how changes in energy density can lead to similar qualitative
behaviors without undermining the core conclusion regarding ghost
star existence.

Moreover, we have included graphical representations of energy
density across our models I and II. These visualizations not only
highlight regions of negative energy density but also facilitate
comparisons between different scenarios. By extending this analysis
to encompass additional models, we can further validate that similar
patterns emerge across various configurations, reinforcing the
universality of our results about ghost stars. The concept of ghost
stars finds support in analogous studies involving dark matter
interactions and other non-standard compact objects, such as boson
stars. By referencing these established theories, we strengthen our
argument for the qualitative universality of ghost stars as viable
astrophysical entities. Finally, we propose that future
observational campaigns targeting gravitational waves or
electromagnetic signals from potential ghost star candidates could
provide empirical validation for our theoretical predictions. If
these observations yield consistent results across different
scenarios involving ghost stars, it would further substantiate the
universality of these findings.

\section{Ghost Star Models admitting Null Complexity Factor}

One of the recently intriguing topics among researchers is the
complexity of anisotropic self-gravitating bodies. Multiple attempts
have been made to develop different techniques that help to assess
the criteria for computing complexity in various fields. Herrera
\cite{37g} contributed to this effort by defining complexity in
terms of all physical factors for a static spherical structure. He
essentially remodeled the notion of Luis Bel by using the orthogonal
splitting of the Riemann tensor. The outcomes of this approach are
certain scalar functions, also known as structure scalars. The
underlying concept is that the system with isotropic pressure and
homogeneous energy density is considered not to be complex.
Therefore, for an inhomogeneous anisotropic configuration, the
factor containing the maximum information is called the complexity
factor. A very recent extension of this definition has been proposed
in the charged framework \cite{1rf}. The Riemann tensor is expressed
in terms of some other factors as
\begin{equation}\label{39a}
R^{\phi\zeta}_{\chi\theta}=C^{\phi\zeta}_{\chi\theta}+16\pi\big(
T^{[\phi}_{[\chi}\delta^{\zeta]}_{\theta]}
+E^{[\phi}_{[\chi}\delta^{\zeta]}_{\theta]}\big)+8\pi
T\left(\frac{1}{3}\delta^{\phi}_{[\chi}\delta^{\zeta}_{\theta]}
-\delta^{[\phi}_{[\chi}\delta^{\zeta]}_{\theta]}\right).
\end{equation}
We now define the tensor $\mathcal{Y}_{\phi\chi}$ in terms of the
Riemann tensor as follows
\begin{equation}\label{39ab}
\mathcal{Y}_{\phi\chi}=R_{\phi\zeta\chi\theta}\textsf{V}^{\zeta}\textsf{V}^{\theta},
\end{equation}
whose alternative form is
\begin{equation}\label{39ac}
\mathcal{Y}_{\phi\chi}=\frac{\mathcal{Y}_T
h_{\phi\chi}}{3}+\bigg(\mathcal{S}_{\phi}\mathcal{S}_{\chi}+\frac{
h_{\phi\chi}}{3}\bigg)\mathcal{Y}_{TF}.
\end{equation}
Equations \eqref{39a}-\eqref{39ac} produce the complexity factor
after some lengthy calculations (see \cite{37g} for more details) as
\begin{equation}\label{45}
\mathcal{Y}_{TF}=\mathcal{E}+\frac{\Pi}{2}-\beta^2r^2.
\end{equation}
In the above equation, the term $\mathcal{E}$ refers to the Weyl
scalar. Its value is expressed as
\begin{equation}\label{45a}
\mathcal{E}=\frac{e^{-\eta}}{4}\left[\lambda''+\frac{\lambda'^2-\eta'\lambda'}{2}
-\frac{\lambda'-\eta'}{r}+\frac{2(1-e^{\eta})}{r^2}\right].
\end{equation}
The zero complexity factor constraint plays a crucial role in
formulating physically feasible models. In the current scenario of
the charged matter configuration, this condition turns out to be
\begin{equation}\label{46}
\Pi=\frac{3\beta^2r^2}{2}+\frac{4\pi}{r^3}\int^{r}_{0}\bigg(\rho+\frac{\beta^2r^2}{8\pi}\bigg)'dr.
\end{equation}
Now, we construct a differential equation in terms of $z$ and $m$
which is derived by utilizing the field equations
\eqref{15}-\eqref{17} together with the above equation as
\begin{align}\nonumber
&-4 r^2 z m'+2 r m'-8 r^2 m z'-8 r^2 m z^2+20 r m z-18 m+4 \beta ^2
r^7 z'\\\label{47}&+4 \beta ^2 r^7 z^2-12 \beta ^2 r^6 z+3 \beta ^2
r^5+4 r^3 z'+4 r^3 z^2-8 r^2 z+8 r=0.
\end{align}
The above condition is a significant aspect when assessing the
stability of stellar configurations. In the following subsections,
we shall use this and establish that the energy density can assume
negative values in certain regions of spacetime. This characteristic
is crucial as it suggests that these ghost stars could be stable
under specific conditions, particularly when they are charged and
exhibit anisotropic fluid configurations. The analysis of the
complexity factor shall reveal that these configurations do not
develop singularities or instabilities that would typically plague
more conventional stellar models, thus allowing for a stable
existence within the framework we propose. Moreover, by applying the
vanishing complexity constraint, we shall derive solutions that
maintain a balance between gravitational forces and the exotic
properties of ghost stars. This balance is essential for ensuring
that the star does not collapse under its own gravity or undergo
catastrophic changes in its structure. These models may indicate
that ghost stars can exist stably due to their unique interplay
between electromagnetic fields and gravitational dynamics, which
allows them to resist collapse while maintaining their exotic
characteristics. Additionally, the implications of these findings
extend beyond mere stability. The ghost stars' ability to exist with
negative energy densities suggests potential avenues for
understanding dark matter and other exotic forms of matter in
astrophysical contexts.

\subsection{Model III}

This model is studied with a specific form of the energy density
which is given by
\begin{equation}\label{48}
8\pi\rho+\beta^2
r^2=\frac{1}{r^2}\bigg[1-9\bigg(\frac{r}{r_{\Sigma}}\bigg)^8\bigg].
\end{equation}
This form is further utilized with Eq.\eqref{15} to produce $m$ as
\begin{equation}\label{49}
m=\frac{r}{2}-\frac{r^9}{2 r_{\Sigma}^8}+\frac{\beta^2 r^5}{2}.
\end{equation}
The motivation for choosing such specific form is that it helps in
the integration of required functions. Moreover, we integrate
Eq.\eqref{47} for $z$ after inserting the above calculated value of
$m$. We are left with
\begin{align}\nonumber
z&=\bigg[-c_1 \bigg\{\frac{1}{1+\frac{\sqrt[3]{-1}}{4}+\frac{1}{4}
(-1)^{2/3}}\bigg((-1)^{\frac{1}{3}+\frac{\sqrt[3]{-1}}{4}}
\bigg(\frac{1}{4}+\frac{\sqrt[3]{-1}}{4}\bigg) r_{\Sigma} ^{-8-2
\sqrt[3]{-1}} \beta ^{-2-\frac{\sqrt[3]{-1}}{2}}
\\\nonumber&\times r^{3+\sqrt[3]{-1}}
\,_2F_1\bigg(1+\frac{\sqrt[3]{-1}}{4},\frac{5}{4}+\frac{\sqrt[3]{-1}}{4};2+\frac{\sqrt[3]{-1}}{4}+\frac{1}{4}
(-1)^{2/3};\frac{r^4}{r_{\Sigma} ^8 \beta
^2}\bigg)\bigg)\\\nonumber&+(-1)^{\frac{1}{3}+\frac{\sqrt[3]{-1}}{4}}
r_{\Sigma} ^{-2 \sqrt[3]{-1}} \beta ^{-\frac{\sqrt[3]{-1}}{2}}
r^{\sqrt[3]{-1}-1} \,
_2F_1\bigg(\frac{\sqrt[3]{-1}}{4},\frac{1}{4}+\frac{\sqrt[3]{-1}}{4};1+\frac{\sqrt[3]{-1}}{4}
\\\nonumber&+\frac{1}{4}
(-1)^{2/3};\frac{r^4}{r_{\Sigma} ^8 \beta
^2}\bigg)\bigg\}+\frac{1}{1-\frac{\sqrt[3]{-1}}{4}-\frac{1}{4}
(-1)^{2/3}}\bigg\{(-1)^{\frac{2}{3}-\frac{1}{4} (-1)^{2/3}}
\bigg(\frac{1}{4}-\frac{1}{4} (-1)^{2/3}\bigg)\\\nonumber&\times
r_{\Sigma} ^{2 (-1)^{2/3}-8} \beta ^{\frac{1}{2} (-1)^{2/3}-2}
r^{3-(-1)^{2/3}} \, _2F_1\bigg(1-\frac{1}{4}
(-1)^{2/3},\frac{5}{4}-\frac{1}{4}
(-1)^{2/3};2\\\nonumber&-\frac{\sqrt[3]{-1}}{4}-\frac{1}{4}
(-1)^{2/3};\frac{r^4}{r_{\Sigma} ^8 \beta
^2}\bigg)\bigg\}+(-1)^{\frac{2}{3}-\frac{1}{4} (-1)^{2/3}}
r_{\Sigma} ^{2 (-1)^{2/3}} \beta ^{\frac{1}{2} (-1)^{2/3}}
r^{-1-(-1)^{2/3}}\\\nonumber&\times \, _2F_1\bigg(-\frac{1}{4}
(-1)^{2/3},\frac{1}{4}-\frac{1}{4}
(-1)^{2/3};1-\frac{\sqrt[3]{-1}}{4}-\frac{1}{4}
(-1)^{2/3};\frac{r^4}{r_{\Sigma} ^8 \beta
^2}\bigg)\bigg]\\\nonumber&\times\bigg\{-(-1)^{\frac{\sqrt[3]{-1}}{4}}
r_{\Sigma} ^{-2 \sqrt[3]{-1}} \beta ^{-\frac{\sqrt[3]{-1}}{2}} c_1
r^{\sqrt[3]{-1}} \,
_2F_1\bigg(\frac{\sqrt[3]{-1}}{4},\frac{1}{4}+\frac{\sqrt[3]{-1}}{4};1+\frac{\sqrt[3]{-1}}{4}\\\nonumber&+\frac{1}{4}
(-1)^{2/3};\frac{r^4}{r_{\Sigma} ^8 \beta
^2}\bigg)-(-1)^{-\frac{1}{4} (-1)^{2/3}} r_{\Sigma} ^{2 (-1)^{2/3}}
\beta ^{\frac{1}{2} (-1)^{2/3}} r^{-(-1)^{2/3}}
\\\label{50}&\times\, _2F_1\bigg(-\frac{1}{4}
(-1)^{2/3},\frac{1}{4}-\frac{1}{4}
(-1)^{2/3};1-\frac{\sqrt[3]{-1}}{4}-\frac{1}{4}
(-1)^{2/3};\frac{r^4}{r_{\Sigma} ^8 \beta ^2}\bigg)\bigg\}^{-1},
\end{align}
where $c_1$ specifies the constant of integration. The expressions
for the radial and tangential pressures are much lengthy and
therefore, we do not provide them here. Figure \textbf{3} presents
the density \eqref{48} within the interior of our developed model
whose negative profile makes it obvious candidate of a ghost star.
The ranges are demonstrated by the following specific values.
\begin{itemize}
\item For $\beta=-0.02$, the energy density is noticed to be
negative for $r>1.482$.
\item For $\beta=-0.04$, this becomes negative for $r>1.381$.
\item For $\beta=-0.06$, we find the energy density to be negative for
$r>1.232$.
\item For $\beta=-0.08$, negative profile of this fluid parameter is observed for
$r>1.093$.
\end{itemize}
\begin{figure}[h!]\center
\epsfig{file=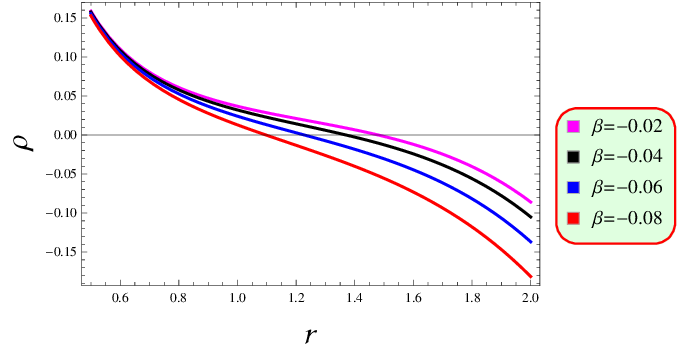,width=0.5\linewidth} \caption{Energy density
\eqref{48} versus $r$ for model III.}
\end{figure}

\subsection{Model IV}

The covariant divergence of the matter source in the charged
framework, i.e., $\nabla^{\phi}(T_{\phi\chi}+E_{\phi\chi})=0$, leads
to the following equation
\begin{align}\label{11}
\mathcal{P}_{r}'=-\frac{\lambda'}{2}(\mu+\mathcal{P}_{r})+\frac{2}{r}
\bigg(\mathcal{P}_{\perp}-\mathcal{P}_{r}+\frac{\beta^2r^2}{4\pi
}\bigg)+\bigg(\frac{\beta^2r^2}{4\pi }\bigg)',
\end{align}
which helps in exploring the hydrostatic equilibrium of the
considered interior geometry. The contribution of the mass function
\eqref{63} along with the field equation \eqref{9} leads to
formulate the metric coefficient as follows
\begin{align}\label{12}
\lambda'=\frac{2(4\pi
r^3\mathcal{P}_{r}-\beta^2r^5+m)}{\beta^2r^6+r(r-2m)}.
\end{align}
Inserting this back into Eq.\eqref{11}, we get
\begin{align}\label{13}
\mathcal{P}_{r}'=-\frac{4\pi
r^3\mathcal{P}_{r}-\beta^2r^5+m}{\beta^2r^6+r(r-2m)}(\mu+\mathcal{P}_{r})+\frac{2}{r}
\bigg(\mathcal{P}_{\perp}-\mathcal{P}_{r}+\frac{\beta^2r^2}{4\pi}\bigg)
+\bigg(\frac{\beta^2r^2}{4\pi}\bigg)'.
\end{align}
An alternative way to examine the mass configuration of a
spherically symmetric structure has been derived by Tolman \cite{42}
and named the Tolman mass, given as
\begin{equation}\label{51a}
m_T=4\pi\int_0^r\bar{r}^2e^{(\lambda+\eta)/2}(T^0_0+E^0_0-T^1_1-E^1_1
-2T^2_2-2E^2_2)d\bar{r}.
\end{equation}
The relationship between the two masses (provided by Misner-Sharp
and Tolman) is presented as follows
\begin{equation}\label{51}
m_T=e^{\frac{\lambda+\eta}{2}}\bigg(m+4\pi
r^3\mathcal{P}_r-\beta^2r^5\bigg),
\end{equation}
and for the vanishing Tolman mass ($m_T=0$), we have
\begin{equation}\label{52}
m+4\pi r^3\mathcal{P}_r-\beta^2r^5=0.
\end{equation}
After inserting this condition in Eq.\eqref{13}, we obtain the
following differential equation consisting of the term $\beta$ that
indicates the impact of charge as
\begin{equation}\label{53}
\mathcal{P}_r^{'}-\frac{3\beta^2r}{4\pi}+\frac{\Pi}{4\pi r}=0,
\end{equation}
which after some manipulation using Eq.\eqref{46} can be written as
\begin{equation}\label{54}
\mathcal{P}_r^{''}+\frac{\rho'}{r}+\frac{4\mathcal{P}_r^{'}}{r}-\frac{13\beta^2}{8\pi}=0.
\end{equation}

Now, the solution of this equation can be obtained by splitting it
further into two equations by
\begin{equation}\label{55}
\mathcal{P}_r^{''}+\frac{3\mathcal{P}_r^{'}}{r}-\frac{13\beta^2}{8\pi}=0,
\end{equation}
\begin{equation}\label{56}
\frac{\rho'}{r}+\frac{\mathcal{P}_r^{'}}{r}=0.
\end{equation}
The solution to the second order differential equation \eqref{55}
for $\mathcal{P}_r$ is calculated as
\begin{equation}\label{57}
\mathcal{P}_r=\frac{c_2}{2}\bigg(\frac{1}{r_{\Sigma}^2}
-\frac{1}{r^2}\bigg)+\frac{13\beta^2}{64\pi}\big(r^2-r_{\Sigma}^2\big),
\end{equation}
where $c_2$ is the constant of integration. Furthermore, we insert
this value in Eq.\eqref{56} to compute the solution for $\rho$ given
as
\begin{equation}\label{58}
\rho=\frac{c_2}{2}\bigg(\frac{1}{r^2}-\frac{1}{r_{\Sigma}^2}
\bigg)+\frac{13\beta^2}{64\pi}\big(r_{\Sigma}^2-r^2\big).
\end{equation}
Feeding back the value of $\mathcal{P}_r$ in Eq.\eqref{52} yields
the mass function
\begin{equation}\label{59}
m=2 \pi  c_2 r \bigg(1-\frac{r^2}{r_{\Sigma} ^2}\bigg)+\frac{1}{16}
\beta ^2 r^3 \big(13 r_{\Sigma} ^2+3 r^2\big).
\end{equation}
Finally, the value of $\Pi$ is easily acquired by inserting
\eqref{58} into \eqref{46} as
\begin{equation}\label{60}
\Pi=\frac{11 \beta ^2 r^2}{8}-\frac{4 \pi  c_2}{r^2}.
\end{equation}

In Figure \textbf{4}, the energy density \eqref{58} is plotted,
showing negative trend for some values of $r$. We define this range
in the following.
\begin{itemize}
\item For $\beta=-0.02$ and $c_2=-0.1$, the energy density becomes negative for $r<1.362$.
\item For $\beta=-0.04$, we find this to be negative for
$r<1.042$.
\item For $\beta=-0.06$, negative behavior of the energy density is
observed for $r<0.698$.
\item For $\beta=-0.08$, this fluid parameter is noticed to be
negative for $r<0.517$.
\end{itemize}
\begin{figure}\center
\epsfig{file=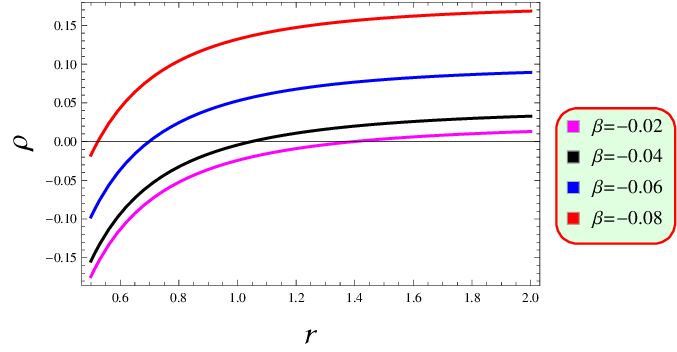,width=0.5\linewidth} \caption{Energy density
\eqref{58} versus $r$ for model IV.}
\end{figure}

\section{Challenges and Perspectives on Ghost Star Existence}

The implications of negative energy density, which, in the context
of quantum field theory, suggests an equation of state with the
parameter $\omega<-1$. This condition can lead to instabilities and
challenges in defining a vacuum state. However, it is crucial to
emphasize that the current study operates primarily within the
classical realm of GR and does not directly invoke quantum
mechanical effects. The models we proposed are grounded in classical
physics, specifically within the Einstein-Maxwell framework. The
idea of ghost stars, as introduced by Zeldovich and Novikov, posits
that these objects can exist under certain conditions where the
energy density is negative in specific regions of spacetime. This
does not contradict classical GR; rather, it suggests that ghost
stars can be treated as theoretical constructs that challenge
conventional notions of compact objects.

Regarding the concern about mass and self-gravitation, it is
important to note that our models demonstrate how ghost stars can
maintain stability despite having regions of negative energy
density. While it is true that if the mass is too small, a
self-gravitating system may not form a compact object, our
formulations account for configurations where the effective
gravitational influence remains significant. The presence of charge
and anisotropic pressures allows for a balance that can lead to
stable configurations even when some regions exhibit negative energy
densities. The interplay between these factors creates scenarios
where ghost stars could manifest as unique astrophysical entities
distinct from traditional stellar structures.

The notion that ghost stars may resemble a ``stellar gas'' rather
than a compact object is an interesting perspective. However, we
argue that our models provide a framework where these objects could
exhibit properties similar to compact stars under specific
conditions. The combination of charge, anisotropic fluid dynamics,
and gravitational effects allows for configurations that challenge
existing paradigms in astrophysics. Winding up this discussion,
while the concerns regarding negative energy density and mass
limitations are theoretically sound, our work aims to explore these
unconventional configurations within a classical framework. The
existence of ghost stars challenges existing paradigms and opens new
avenues for understanding stellar evolution and dark matter
interactions. We believe that further exploration of these models
could yield significant insights into both theoretical physics and
observational astrophysics.

\section{Conclusions and Future Directions}

Investigating the potential existence of spherical objects with
negligible total mass, four novel solutions to the field equations
have been formulated in this paper within the context of an
electromagnetic field. For these types of models (referred to ghost
stars), the energy density must be negative, not everywhere, but in
some regions of the considered fluid distribution. The local
anisotropy has been considered in the self-gravitating interior
geometry and the corresponding field equations have been developed.
After this, we have rewritten the field equations in terms of the
interior mass and a particular generating function. The fundamental
forms of the matching criteria have also been utilized to find a
particular relation between the considered generating function and
the spherical boundary. The explanation of how we have developed
models I and II is given as follows.
\begin{itemize}
\item Model I has been developed using the conformally flatness condition and a
specific density profile. We have observed the energy density for
this model to be negative (Figure \textbf{1}).
\item We have then formulated model II by taking the above former
condition along with Gokhroo and Mehra ansatz into account. The
corresponding energy density remains negative for some values of $r$
(Figure \textbf{2}).
\end{itemize}
Furthermore, the highlights regarding the remaining two solutions
are presented in the following.
\begin{itemize}
\item The vanishing complexity condition along with a known profile
of the energy density has been utilized to develop our model III.
This confirms the negative density within the specific range (Figure
\textbf{3}).
\item To formulate model IV, we have established the Tolman-Oppenheimer-Volkof
equation, a sum of different forces (such as gravitational,
electromagnetic, etc.) which in principle explains the equilibrium
scenario of a celestial object. Further, the above-mentioned former
constraint and disappearing active gravitational mass have been
considered. The corresponding density observes to be negative for
the certain range (Figure \textbf{4}).
\end{itemize}
Table \textbf{1} presents a detailed overview on the numerical range
of the radial coordinate in which the energy density corresponding
to all our resulting models becomes negative.
\begin{table}
\scriptsize \centering \caption{Range of the radial coordinate for
which the energy density is negative.} \label{Table1}
\vspace{+0.07in} \setlength{\tabcolsep}{1.8em}
\begin{tabular}{cccccc}
\hline\hline $\mathbf{Values~of~\beta}$ & \textbf{Model I} &
\textbf{Model II} & \textbf{Model III} & \textbf{Model IV}
\\\hline
$\mathbf{-0.02}$ & $r<$ 0.634 & $r>$ 0.956 & $r>$ 1.482 & $r<$ 1.362
\\\hline
$\mathbf{-0.04}$ & $r<$ 0.575 & $r>$ 0.718 & $r>$ 1.381 & $r<$ 1.042
\\\hline
$\mathbf{-0.06}$ & $r<$ 0.544 & $r>$ 0.452 & $r>$ 1.232 & $r<$ 0.698
\\\hline
$\mathbf{-0.08}$ & $r<$ 0.516 & $r>$ 0.339 & $r>$ 1.093 & $r<$ 0.517 \\
\hline\hline
\end{tabular}
\end{table}

The primary difference between \cite{1r} and the current analysis
lies in our incorporation of electromagnetic interactions, which
fundamentally alters the dynamics of ghost stars. In standard GR
models, ghost stars are characterized by negative energy density
regions, leading to intriguing properties such as vanishing total
mass. However, these models do not account for the influence of
charge or electromagnetic fields on the structure and stability of
such objects. By introducing the electric charge, we have
investigated how these additional forces affect the energy density
and overall configuration of ghost stars. Since our study expanded
upon existing models by analyzing charged anisotropic fluid
configurations, this allows us to explore a broader spectrum of
physical scenarios and provides insights into how electromagnetic
interactions might stabilize or destabilize these exotic objects.
The field equations reveal intricate relationships between mass,
charge, and energy density that were previously unexplored. It is
important to mention here that our results are in-line with those
obtained without considering the impact of an electromagnetic field
\cite{1r}. Also, the present results reduce to \cite{1r} when
substituting $q=0$ or $\beta=0$.

The formation of a ghost star is inherently tied to the processes
that occur during gravitational collapse. It is essential to
recognize that this collapse is often accompanied by significant
radiative phenomena. Various studies have explored the efficiency of
energy release during such collapses, leading to differing
conclusions. Some researchers suggested that a complete energy
release, where all mass is converted into radiation, is feasible
under certain conditions \cite{1sa,1sb}. Conversely, other experts
argued that achieving 100$\%$ efficiency is unlikely when
considering realistic physical constraints, particularly the
requirement for positive energy density \cite{1sc}. The implications
of violating this energy condition, as seen in our models, lend
credence to the possibility of complete efficiency being attainable.
Consequently, if a substantial emission of radiation is detected, it
could serve as an indicator of a ghost star's presence.

Some remarks must be stressed here that encourage researchers to
extend this work in different directions in the future.
\begin{itemize}
\item The ghost stars have been discussed in GR with and without
including the effects of charge. This work should now be extended in
modified theories to see whether such kind of fluid distributions
could exist or not.
\item It has been observed that the collapsing objects can be
discussed through shadows as depicted by recent observations
\cite{1s}-\cite{1v}. From this perspective, could it be possible to
explore the existence of ghost stars through their shadows?
\item In the present study, we have formulated different models
by assuming anisotropic matter content. It becomes clear that the
same models can be discussed under the consideration of an isotropic
fluid, and thus, their existence should be evaluated.
\item Exploring the potential outcomes of an initial ghost star
transitioning into a object with positive mass through radiation
absorption could yield valuable insights. While this concept (of a
compact entity absorbing radiation) may seem unusual, it has
previously been proposed as a mechanism for explaining the gas
origins in quasars \cite{1va}. A semi-numerical representation of
such a theoretical framework can be found in \cite{1vb}. \\
\end{itemize}
\textbf{Data Availability Statement:} This manuscript has no
associated data.

\end{document}